\documentclass[ajp,twocolumn,showpacs]{revtex4}
\usepackage{graphicx}

\newcommand{\be}{\begin{equation}}
\newcommand{\ee}{\end{equation}}
\newcommand{\beqa}{\begin{eqnarray}}
\newcommand{\eeqa}{\end{eqnarray}}

 \newcommand{\lan}{\langle}
  \newcommand{\ran}{\rangle}
 
\begin{document}
  
  \title{ Estimating errors reliably in Monte Carlo simulations of the Ehrenfest model}
 \author{Vinay Ambegaokar}
\affiliation{Laboratory of Atomic and Solid State Physics, Cornell University, Ithaca, New York 14853, USA}

\author{Matthias Troyer}
\affiliation{Theoretische Physik, ETH Zurich, 8093 Zurich, Switzerland}

\date{June 3, 2009}

 \begin{abstract}
Using the Ehrenfest urn model we illustrate the subtleties of error estimation in Monte Carlo simulations. We discuss how the smooth results of correlated sampling in Markov chains can fool one's perception of the accuracy of the data, and show (via numerical and analytical methods) how to obtain reliable error estimates from correlated samples.
  \end{abstract}
   \maketitle
\section{Introduction and Summary}

The Ehrenfest urn model \cite{ PTE}, is sometimes picturesquely described as fleas jumping between dogs.  One imagines a sub-system of $N$ numbered fleas residing on dog A or dog B, each jumping from one dog to the other when its number is called.  This model has been previously used  \cite{MK, Em}, including in this journal \cite{AC}, to illuminate thermodynamic equilibration and equilibrium.  Monte Carlo simulations of the process are particularly instructive.  One of us noticed \cite{RAL}, but let pass without investigation, the errors associated with such calculations. This neglect is remedied in the present paper, raising issues known to specialists but perhaps not widely enough appreciated.  The tutorial exposition given here may therefore be of general interest. 

It is worth stressing that, although the message of this paper is that single flea hops are an inefficient way to sample the steady-state, the process is ideally suited to understanding thermodynamically irreversible transitions from unlikely to likely configurations \cite{MK,Em,AC,RAL}, as well as fluctuations in equilibrium, which in typical physical situations also proceed in small steps.

The paper is organized as follows.  Section 2 contains a brief description of the essence of the Monte Carlo method.  Although the procedure is useful in cases where an enumeration  of possibilities is prohibitively difficult, the urn model is simple enough to allow explicit analysis.  In the main body of the paper we exploit only the fact that the steady-state probability of $n$ fleas on dog A is a binomial distribution, and use this as a check for various numerical simulations.  In section 3, we  show that trials of $N$-flea configurations yield good results with expected errors.  We then simulate the single flea transfer used in refs.\cite{Em,AC, RAL} and encounter the apparent inaccuracies mentioned above.  In section 4, correlations between successive samples and their effect in reducing the number of independent trials is studied, and a numerical method (``binning analysis") is used to illuminate and eliminate the problem, leading to the conclusions of Section 5.  In an Appendix, the Markov (i.e. memoryless) random process underlying single flea transfers is treated analytically, using methods similar to those in ref.\cite{MK}, revealing nice features of the approach to equilibrium and the autocorrelation problem.  

\section{The Monte  Carlo method}

It is told that Stanislav Ulam \cite{Ulam} invented the Monte Carlo method in the 1940s when playing Solitaire while lying sick in bed. He wanted to know the probability of winning in Solitaire but was faced with the problem that with $52! \approx 10^{68}$ different ways of arranging the cards  he could never exactly calculate the chance of winning. He realized, however, that by just playing 100 games and counting the number of wins he could already get a pretty good estimate. 

This insight suggested a way of tackling the problem caused by the exponential growth with size in the number of states of a statistical system.  In a general statistical context, one might wish to calculate weighted averages over configurations.  However, even in our very simple model the number of ways of distributing fleas between dogs is $2^{N}$.   These configurations may be enumerated by $2^N$ $N$-dimensional vectors  $\vec{x}$ of which each element  $x_n$, $1 \le n \le N$, can take on two values.  If each configuration is assigned a normalized weight $p(\vec {x})$,  $\sum_{\vec{x}}p(\vec{x}) = 1$, the weighted mean of an arbitrary function of the configuration, $A(\vec{x})$ is
\be 
\langle A \rangle \equiv \sum_{\vec{x}} A(\vec{x}) p(\vec{x}). \label {eq:sum} 
\ee
An exact summation over all states is, in general, impossible for $N > 40$, even on the most powerful supercomputers.  The Monte Carlo method  \cite{MonteCarlo}, which Ulam named after the famous casinos in Monaco \cite{LasVegas}, tries to estimate such sums by a partial sum over a sample of only $M \ll 2^N$ configurations $\vec{x}_i$ 
\begin{equation}
  \overline{A} \equiv  \frac{1}{M}\sum_{i=1}^M A_i, \label {eq:sample}
\end{equation}
where the configurations $\vec{x}_i$ are chosen randomly with the correct probability $p(\vec{x})$, and we have introduced the shorthand notation $A_i \equiv A(\vec{x}_i)$.
 
Choosing the sample randomly and with the correct probabilities is as crucial here as in opinion polls before presidential elections: only a truly random and representative sample will give meaningful results.

The estimate $\overline A$ of the true expectation value $\langle A \rangle$ is a fluctuating quantity that  will deviate from the true value. According to the central limit theorem, $\overline{A}$ is normally distributed around  $\langle A \rangle$ with a standard deviation $\Delta_A$ that we shall calculate below. 

As a warmup let us show that the expectation value of  $\overline A$ is indeed $\langle A \rangle$:
\begin{eqnarray}
\langle \overline{A} \rangle &=&  \langle \frac{1}{M} \sum_{i=1}^M A_i \rangle \nonumber \\
&=&  \frac{1}{M} \sum_{i=1}^M  \langle A_i \rangle \\
&=&  \frac{1}{M} \sum_{i=1}^M  \langle A \rangle = \langle A  \rangle \nonumber.
\end{eqnarray}
In going from the first to the second line we have used linearity of the expectation value; going from the second to the third line we have made use of the fact that the samples $\vec{x}_i$ are all chosen from the same distribution $p(\vec{x})$, so that the $A_i$s have the expectation value given by Eq.~(\ref{eq:sum}).

Similar reasoning allows the calculation of the average of the square of the sample mean.

\begin{eqnarray}
 \left\langle{\overline{A} ~}^2\right\rangle&=&\left\langle \left(\frac{1}{M}\sum_{i=1}^M  A_i \right)^2 \right\rangle 
 = \frac{1}{M^2} \sum_{i=1}^M\sum_{j=1}^M \langle A_iA_j \rangle\nonumber\\ 
 &= &\frac{1}{M^2} \sum_{i=1}^M  \langle A_i^2 \rangle 
+\frac{M-1}{M} \langle A \rangle^2 \nonumber\\ 
&=& \frac{1}{M}\langle A^2 \rangle + \frac{M-1}{M} \langle A \rangle^2,\label{eq:fluct}
\end{eqnarray}
where we have inserted the definition of the average (\ref{eq:sample}), used the linearity of the expectation value,  and also exploited the fact that for independent samples $\vec{x}_i$ and $\vec{x}_j$ the expectation value for $i \ne j$ factorizes as
\begin{equation} 
\langle A_iA_j \rangle = \langle A_i\rangle \langle A_j \rangle 
= \langle A \rangle^2 .\label{eq:independent}
\end{equation}
The statistical error $\Delta_A$, the root-mean-square deviation of the sample mean $\overline{A}$ from the true expectation value $\langle A \rangle$, is thus given by 
 \begin{eqnarray}
 \Delta_A^2 &\equiv& \left\langle \left(\overline{A}-\langle A \rangle\right)^2\right\rangle \nonumber \\
&=& \frac{1}{M^2} \sum_{i=1}^M \langle A_i^2 \rangle 
-  \frac{1}{M} \langle A \rangle^2 \nonumber \\
&=&\frac{1}{M}\left( \langle A ^2 \rangle - \langle A \rangle^2\right) \nonumber \\
&\equiv& \frac{1}{M}{\rm Var}A,\label{eq:error1}
\end{eqnarray}
which is the basis of the central limit theorem. It is, however, more useful to express the error in terms of the sampled $A_i$s. A na\"ive guess would be to estimate the variance as $\overline{A^2} -\overline{A}^2$, where
\be
\overline{A^2}\equiv\frac{1}{M}\sum_{i=1}^M A_i ^2.
\ee
Calculating the expectation values via Eq.~(\ref{eq:fluct}) shows that
\begin{equation}
\left\langle \overline{A^2} -\overline{A}^2\right \rangle = \frac{M-1}{M}{\rm Var}A.
\end{equation}
The true estimator is thus 
\begin{equation}
{\rm Var}A \approx \frac{M}{M-1}\left(\overline{A^2} -\overline{A}^2\right),\label{eq:var}
\end{equation}
where the (small) fluctuations of the right hand side of Eq.~(\ref{eq:var}) have been ignored.
Taking the square root, we obtain the final result
\begin{equation}
\Delta_A =\sqrt{\frac{{\rm Var}A}{M}} \approx \sqrt{\frac{\overline{A^2} -\overline{A}^2}{M-1}}. 
 \label{eq:errorn}
\end{equation}
The $-1$ in the denominator, which is of course irrelevant for the large values of $M$ in the numerical simulations below, reflects the loss of one piece of information in calculating the sample mean.

\section{Dogs and Fleas} 
After these preliminaries,  let us consider the fleas on two dogs game as played in references \cite{RAL, AC}.  The game starts with two dogs -- a flea-ridden dog B(urnside) with $N=50$ fleas and a clean dog A(nik).  Once per time step a randomly chosen flea hops from one dog to the other, so that asymptotically the probability of a flea being on one of the dogs is 1/2. In this simple case, it is possible to analytically calculate the probability distribution $P[n]$ for having $n$ of the $N$ fleas on one dog.  It is the binomial distribution
\begin{equation}
P_{eq} [n] = \frac{1}{2^N}\pmatrix {N\cr n}=  \frac{1}{2^N}\frac{N!}{n!(N-n)!}
\label{eq:exact}
\end{equation}
This exact solution will be very useful as a test for our Monte Carlo simulations. 

\subsection{Direct Sampling}

\label{sec:direct}
Our first Monte Carlo simulation will not yet follow the above game, but will directly sample the asymptotic distribution. For each sample, we loop over all fleas and draw a uniformly distributed random binary integer $u\in\{0,1\}$. If $u=0$ the flea is positioned on Anik, otherwise on Burnside. In order to estimate the distribution $P[n]$ for the number of fleas $n$ on Anik it will be sufficient to record a histogram $H[n]$ counting how often $n$ fleas ended up on her. From this histogram we can compute an estimate for $P[n]$ as
\begin{equation}
\overline{P[n]} = \frac{1\cdot H[n]+0\cdot(M-H[n])}{M} = \frac{H[n]}{M},
\end{equation}
since our estimator is $1$ whenever there were $n$ fleas on Anik and $0$ otherwise.
Since $1^2=1$ and $0^2=0$ we get the same estimator for the square
\begin{equation}
\overline{P[n]^2}=\frac{1^2\cdot H[n]+0^2\cdot(M-H[n])}{M} = \frac{H[n]}{M},
\end{equation}
from which we obtain the error estimate
\begin{equation}
\Delta_{P[n]}\approx \sqrt{\frac{H[n]/M-H[n]^2/M^2}{M-1}}
\end{equation}

\begin{figure}[t]
\centerline{\includegraphics[width=8cm]{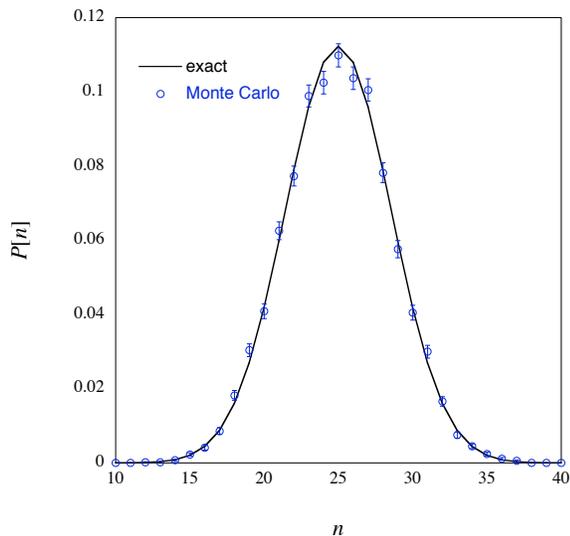}}
\caption{Comparison of the flea distribution $P[n]$ obtained in a direct Monte Carlo simulation with the exact asymptotic result. $M=10,000$ samples were recorded.}
\label{fig:direct}
\end{figure}

In Fig. \ref{fig:direct} we compare the exact solution to the Monte Carlo solution for $M=10,000$ samples and find that, as expected from the normal distribution, the exact solution lies within error bars about 2/3 of the time. The Monte Carlo simulation is working well!

\subsection{The Dogs and Fleas Simulation}
\label{sec:sim}
Next we want to implement the simulation of the dog and fleas game,  Here we will repeat these simulations, observe discrepancies, and explain their origin.

As introduced above, we start with all $N=50$ fleas on Burnside and hence $n=0$. In each simulation step we will then pick one of the $N$ fleas at random, by drawing a uniform integer random number $u$ between $1$ and $N$ and move that flea to the other dog. In practice we label the fleas so that the fleas $1,\ldots,n$ are on Anik and the fleas $n+1,\ldots,N$ on Burnside. Hence if $u\le n$ we move a flea from Anik to Burnside and decrease $n$ by one, otherwise we move a flea in the opposite way and increase $n$ by 1. 

\begin{figure}[tb]
\centerline{\includegraphics[width=8cm]{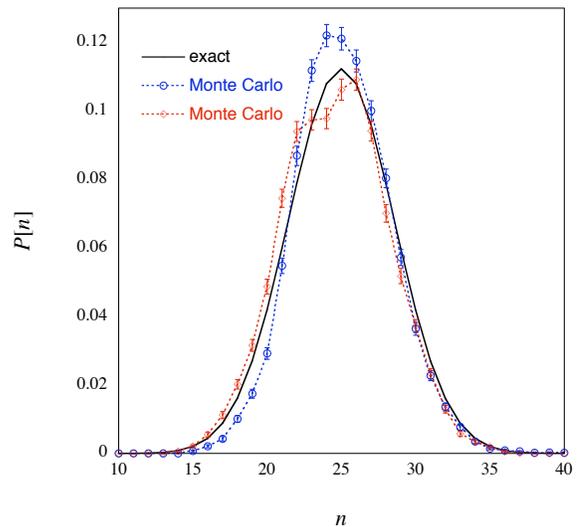}}
\caption{Comparison of the flea distribution $P[n]$ obtained in Monte Carlo simulations of the original dog and fleas game with the exact asymptotic result. Two different random seeds were used, $M=10,000$ and $M/5$ steps were used for equlibration. Something is obviously wrong since the exact results are significantly outside the error bars, not even the two simulations agree, and the asymmetric shape cannot be right.}
\label{fig:fleasim}
\end{figure}

In our simulation we need to wait a while until the fleas have equilibrated and we can expect to observe the asymptotic distribution. We thus perform $M/5$ flea hops for equilibration, without recording any measurements. Only then do we start with the actual simulation and perform $M$ flea hops, recording a histogram $H[n]$.

In simple examples like this simulation we might actually be able to guess the number of steps needed for equilibration. As we show in the appendix, only about 50 hops are needed to reach equilibrium. Why then did we throw away 20\%, or 2,000 samples? The reason is that in more complex cases we often have no idea of the actual equilibration times. It is then strongly recommended to err on the side of throwing away too many samples rather than too few. By throwing away the first 20\% of our samples we increase our statistical error by only about 10\% (remember the inverse square root scaling of the error with the number of samples), which is a small price to pay to be on the safe side regarding equilibration.

In Fig. \ref{fig:fleasim} we again compare to the exact solution and observe deviations remarked on before \cite{RAL}.  At first sight, the deviations are puzzling, since the curves look smooth.  However, the asymmetric shapes cannot be correct, and the errors bars, calculated  using Eq.~(\ref {eq:errorn}) with $M=8000$ are evidently too small.   That these features are general can be seen by repeating the simulations with different random seeds: sometimes the results look mostly right, but often they are just plainly wrong as in Fig. \ref{fig:fleasim}. The large variations observed also confirm that something is wrong with the error estimates. 

A little further thought suggests the reason.  Eq.~(\ref {eq:errorn}) is an estimate for the relative deviation from the mean of $M$ trials of a binomial process with a success probability estimate $\overline{P[n]}$.  But, $M$ single flea hops is not the same as $M$ trials of the whole distribution as performed to obtain Fig. \ref {fig:direct}.

\section{Autocorrelation Effects and Error Estimates}

We need to reconsider the derivation of the errors in equations (\ref{eq:error1}) to (\ref{eq:errorn}). The only assumption, besides a finite variance, was in Eq.~(\ref{eq:independent}): the independence of samples $\vec{x}_i$ and $\vec{x}_j$ for $i\ne j$. While this independence was clearly given in the direct simulation --- at least as long as we use independent random numbers to create the flea distributions --- it 
is not true of the original dogs and fleas simulation, in which subsequent samples differ only by a single random process. They form what is called a ``Markov chain."   As just remarked, this method of sampling evidently explores the space of states much less efficiently than the calculation of Fig. \ref{fig:direct}, in which every flea is addressed at every trial.  Equation (\ref{eq:independent}) and thus also the error estimate (\ref{eq:errorn}) are not valid for correlated samples from a Markov chain.  The correlation between samples is also responsible for the smooth shape of the results, which fools our intuition about the errors of the results.  
  
In the following we will discuss two methods for obtaining  reliable errors of a Monte Carlo simulation

\subsection{Error Estimates from Independent Simulations}

\begin{figure}[tb]
\centerline{\includegraphics[width=8cm]{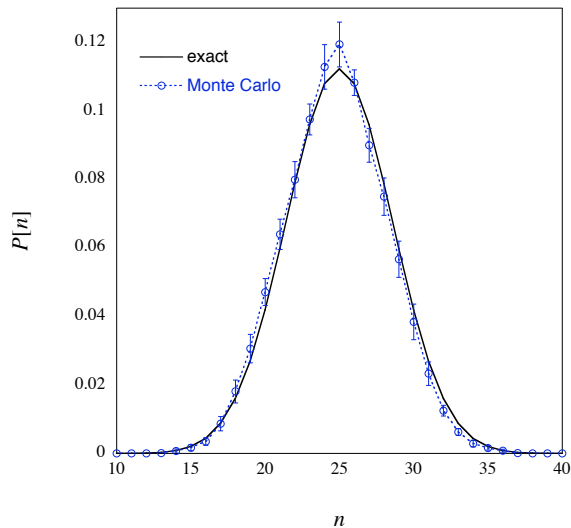}}
\caption{Comparison of the flea distribution $P[n]$ obtained in Monte Carlo simulations of the original dog and fleas game with the exact asymptotic result. Now $L=10$ independent simulations were performed for a total of $M=10,000$ measurements. Each simulation performing $M/L$ measurements was equilibrated for $M/5$ steps. Now the error bars, estimated from the $L=10$ independent simulations are much larger and agree with the exact result, but at the cost of having to equilibrate $L$ simulations.}
\label{fig:fleasindep}
\end{figure}

The easiest way of obtaining reliable error estimates is to create independent samples. To obtain them we perform the simulation multiple times with different random seeds. In each simulation we record an estimate for $P[n]$. Then we obtain a final estimate for $P[n]$ by averaging the $P[n]$ obtained in the individual simulations and an error estimate by applying Eq.~(\ref{eq:errorn}) to the $P[n]$ obtained from these independent simulations. 

We show results from performing $L=10$ simulations of $M/L=1000$ measurements each in Fig. \ref{fig:fleasindep}. While we still see large deviations, and the maximum appears too high as in Ref. \cite{AC, RAL}, the error bars are now much larger and appear correct --- they include the correct value most of the time!

While performing $L$ independent simulations gives reliable error bars we pay the cost that each of the $L$ simulations needs to be equilibrated independently, so that in our case we performed $LM/5=20,000$ equilibration steps in addition to $M=10,000$ measurement steps. 

\subsection{Error Estimates from Uncorrelated Samples}
\label{sec:indep}
\begin{figure}[tb]
\centerline{\includegraphics[width=8cm]{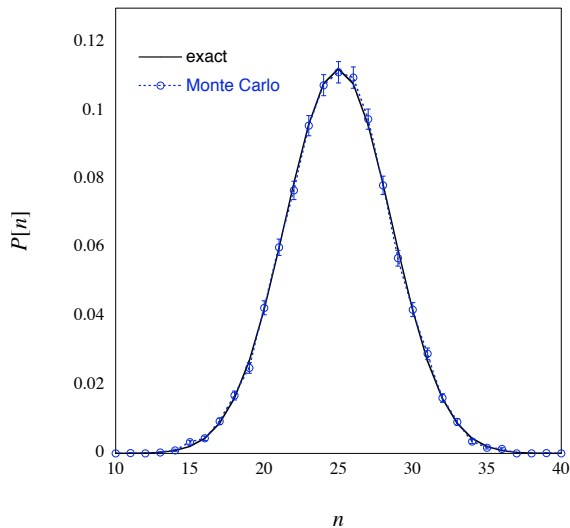}}
\caption{Comparison of the flea distribution $P[n]$ obtained in Monte Carlo simulations of the original dog and fleas game with the exact asymptotic result. Now $N_{\rm hop}=99$ flea hops were made between the $M=10,000$ measurements and $M/5$ steps were used for equlibration. $N_{\rm hop}=99$ seem to be enough hops to decorrelate the samples and give reliable error estimates .}
\label{fig:uncorrelated}
\end{figure}

Another way of obtaining reliable errors is not to measure after every flea hop, but to let many fleas hop before performing a measurement. In Fig. \ref{fig:uncorrelated} we show the results from a simulation performing $N_{\rm hop}=99$ flea hops \cite{ODD} between each of the $M=10,000$ measurements. 

Now the Monte Carlo results agree with the exact results and the error bars are again much smaller, but at the cost of having to perform $N_{\rm hop}=99$ times more flea hops, and also losing all of the potentially useful information between measurements. In addition, we have no way of knowing whether $N_{\rm hop}=99$ hops between measurements are sufficient to create uncorrelated samples for which Eq.~(\ref{eq:independent}) holds, or whether a much smaller suffices or a much larger number is needed. 
\subsection{Error Estimates for Correlated Samples}

While the above approach clearly demonstrates that indeed the correlations between samples $\vec{x}_i$ and $\vec{x}_j$ are the origin of our problems, it is not a viable solution. Instead let us correct the error estimate Eq.~(\ref{eq:errorn}) for the case of correlated samples by including the terms which we omitted above under the assumption of independence (\ref{eq:independent}) to obtain:
\begin{equation}
\Delta_A^2=\frac{{\rm Var}A}{M}+\frac{1}{M^2}\sum_{i\ne j=1}^M\left(\langle A_iA_j \rangle-\langle A\rangle^2\right)
\label{eq:error_corr}
\end{equation}
Previously we had assumed that, due to independence the second term is zero. Let us now replace the assumption of independence by a rapid decay as $|i-j|\rightarrow\infty$ \cite{RC} and rewrite the second term as
\begin{eqnarray}
&&\frac{1}{M^2}\sum_{i\ne j=1}^M\left(\langle A_iA_j \rangle-\langle A\rangle^2\right) \nonumber \\
&=&\frac{2}{M^2}\sum_{i < j=1}^M\left(\langle A_iA_j \rangle-\langle A\rangle^2\right) \nonumber \\
&=&\frac{2}{M^2}\sum_{i=1}^M\sum_{t=1}^{M-i}\left(\langle A_iA_{i+t} \rangle-\langle A\rangle^2\right) \nonumber \\
&=&\frac{2}{M}\sum_{t=1}^{M-1}\left(\langle A_1A_{1+t} \rangle-\langle A\rangle^2\right) \nonumber \\
&\approx&\frac{2}{M}\sum_{t=1}^{\infty}\left(\langle A_1A_{1+t} \rangle-\langle A\rangle^2\right) \nonumber \\
&\equiv&\frac{2}{M}({\rm Var}A)\tau_A. \label{eq:corr}
\end{eqnarray}
 Going from the second to third line we relabeled the indices, in the next line we used the identical distributions to limit the sum over $i$ to the first index, in the fifth line we extended the sum over $t$ to infinity since the correlations are expected to decay fast enough, and in he last line we used the definition of the integrated autocorrelation time $\tau_A$ of $A$:
 \begin{equation}
 \tau_A\equiv \frac{\sum_{t=1}^\infty \left(\langle A_1A_{1+t} \rangle-\langle A\rangle^2\right)}{\langle A^2\rangle -\langle A \rangle^2} \label{eq:tau}
 \label{eq:auto}
 \end{equation}
 Inserting Eq.~(\ref{eq:corr}) into Eq.~(\ref{eq:error_corr}) we end up with the final error estimate
 \begin{equation}
\Delta_A = \sqrt{\frac{{\rm Var}A}{M}(1+2\tau_A)}
\label{eq:errorc}
 \end{equation}
and see that due to correlation effects the error is increased by a factor of $\sqrt{1+2\tau_A}$.  Eq.~(\ref{eq:errorc}), in fact, very nicely gives the effective number of uncorrelated samples as $[M/(1 + 2\tau_A)] <M$.  While this explains the failures of the simple error estimate (\ref{eq:errorn}), it does not help us much yet since the estimation of $\tau_A$ via Eq.~(\ref{eq:tau}) is expensive and cumbersome. A fast and easy way of estimating errors is explained below and an exact calculation of the autocorrelation time for this model is presented in the appendix.

\subsection{Error Estimates from a Binning Analysis}
The binning analysis is a method of analyzing Monte Carlo data based on Eq.~(\ref{eq:errorc}). It provides both an estimate for the error $\Delta_A$ and for the integrated autocorrelation time $\tau_A$.

Starting from the original series of measurements
\begin{equation}
A_{i}^{(0)}=A_i
\end{equation}
we iteratively create ``binned'' series by averaging 
over two consecutive entries:
\begin{equation}
A_{i}^{(l)}:={1\over 
2}\left(A_{2i-1}^{(l-1)}+A_{2i}^{(l-1)}\right)
\end{equation}
for $i=1,\ldots,M_{l}\equiv M/2^{l}$.

Every entry in this new and shorter time series is the average of two adjacent values in the original one. The mean of the new binned time series is the same as the original time series. The averaged values are, however, less correlated than the original ones. The (incorrect) error estimates using the equation (\ref{eq:errorn}) for uncorrelated samples gives errors
\be 
\Delta_A^{(l)} 
 \approx 
\sqrt{\frac{1}{M_{l}(M_l-1)}\sum_{i=1}^{M_{l}}\left(A^{(l)}_{i}-\overline{A^{(l)}}\right)^{2}} ,
\ee
that increase as a function of bin size $2^{l}$. These errors converge to the correct error estimate:
\begin{equation}
\Delta_A = \lim_{l\rightarrow\infty}\Delta_A^{(l)}
\end{equation}
when the bins become uncorrelated for sizes $2^{l}\gg\tau_{A}$.

This binning analysis thus gives a reliable recipe for estimating errors 
and autocorrelation times. One has to calculate the error estimates for 
different bin sizes $l$ and check if they converge to a limiting 
value. If convergence is observed the limit 
$\Delta_A$ is a reliable error estimate, and $\tau_{A}$ can 
be obtained from equation (\ref{eq:errorc}) as
\begin{equation}
\tau_{A}={1\over 2}\left[\left(\frac{\Delta 
_A}{\Delta_A^{(0)}}\right)^{2}-1\right]
\end{equation}

If however no convergence of the $\Delta_A^{(l)}$ is observed we know 
that $\tau_{A}$  is longer than the simulation time and we 
have to perform {\it much} longer simulations to obtain reliable 
error estimates.

\begin{figure}[tb]
\centerline{\includegraphics[width=8cm]{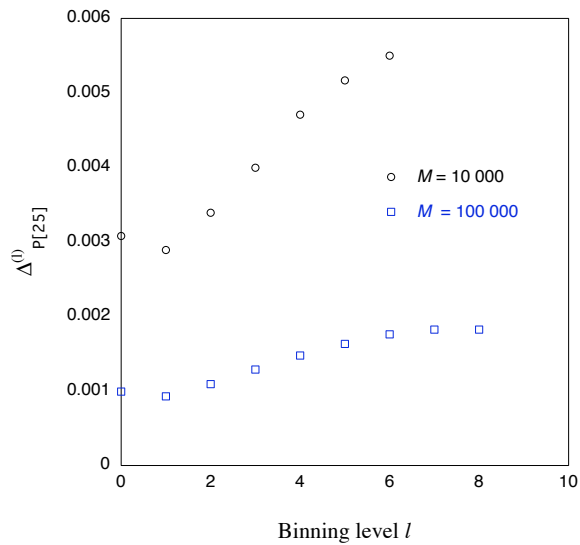}}
\caption{Binning analysis of the error $\Delta_{P[25]}$ of the central value $P[25]$ of the distribution. it is clearly seen that for $M=10,000$ samples the errors have not yet converged, while for $M=100,000$ samples convergence starts to be seen. At least $M=100,000$  samples have to be taken to get reliable results.}
\label{binningfleas}
\end{figure}

\begin{figure}[tb]
\centerline{\includegraphics[width=8cm]{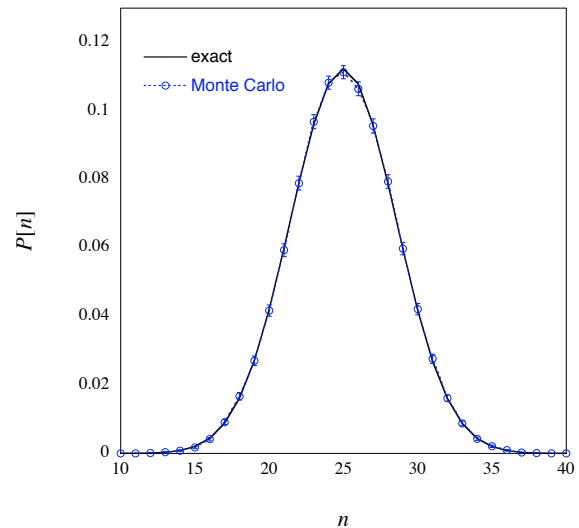}}
\caption{Comparison of the flea distribution $P[n]$ obtained in Monte Carlo simulations of the original dog and fleas game with the exact asymptotic result. This time $M=100,000$ correlated measurements were taken, and the errors calculated using a binning analysis: all is fine!}
\label{fig:correlated}
\end{figure}

Let us redo one of the simulation of section \ref{sec:sim} and perform a binning analysis. In Fig. \ref{fig:correlated} we show our results for $M=100,000$ measurements calculating the errors using the binning analysis. Now everything is in order! 

It is worth noting that the autocorrelation time depends on the variable being sampled.  For example, calculating this quantity for the number $n$ of fleas yields 24.0, which is larger than the value obtained from Fig. \ref{binningfleas} for the peak of the histogram.

To implement the binning analysis it is not necessary to store the full time series. Instead memory of $2\log_2M$ numbers is sufficient. Interested reader are encouraged to look at the implementation in the source file {\tt src/alps/alea/simplebinning.h} of the ALPS libraries \cite{ALPS}. 

\section {Conclusions:  Lessons Learned}

In the discussion of the dogs and fleas simulation we have seen some of the subtleties and pitfalls in estimating reliable errors for results of Monte Carlo simulations. Correlation effects make it necessary to perform a binning analysis instead of using the simple Eq.~(\ref{eq:errorn}) which is valid only for independent samples. 

We have not touched on the issue of cross-correlations between different quantities, that influence error estimates of e.g. the specific heat $c_v=(\langle E^2\rangle - \langle E \rangle^2)/k_BT^2$. To calculate such errors, a bootstrap or jackknife method \cite{NR} is required in addition to a binning analysis.

An important lesson learned is that a reliable analysis of errors of a simulation can be much harder than performing the simulation, but is an essential part for any numerical project. We could have drawn incorrect conclusions and conjectured a new physical phenomenon based on our too small error bars!  

We have also seen that using improved methods, such as the direct sampling of the distribution in Fig.\ref{fig:direct}, smaller errors and more reliable results can be obtained. Unfortunately direct sampling is impossible in all but the simplest models --- but still improved algorithms are the key to reliable large scale simulations.  It is interesting that over the past three decades progress in algorithms for the simulation of the Ising model has outperformed Moore's law:  running modern algorithms on 30 year old computers would be faster than running 30 year old algorithms on the fastest supercomputers of today \cite{TL}!   

All of the programs used to produce the data in this paper are included in the {\tt example/sampling} directory of the latest release of the ALPS libraries \cite{ALPS}.

\section* {Acknowledgements}

VA thanks his colleague Erich Mueller for suggesting the generating function method used in the Appendix, and Cornell graduate students Frank Petruzielo and Bryan Daniels for C++ instruction. MT acknowledges support of the Aspen Center for Physics.

\section*{APPENDIX}
The evolution of the number of fleas on Anik in our Monte Carlo simulation is done probabilistically using a ``Markov process."  Let $P_i[n], n=0,1, \dots N, $ be the $i$-th update of the probability of $n$ fleas on Anik.  Then \cite{MK, RAL, AC}

\beqa 
P_{i+1}[n]&=& {N-n+1\over N} P_i[n-1] + {n+1\over N} P_i[n+1] \nonumber \\
&=&{1\over N}
 \sum_{n'=0}^N M[n,n'] P_i[n'],
\label{eq:mark}
\eeqa
where the coefficients are the relative probabilities for a flea to hop on or off, written in the last line in terms of  a $(N+1)\times(N+1)$ tridiagonal matrix $M$ with the entries $N,N-1, N-2,\ldots 2,1$ on the sub-diagonal, $1, 2, 3\ldots N-1, N$ on the superdiagonal, and zeros elsewhere
\be M \equiv
\pmatrix{0&1&0&\ldots&0&0&0\cr N&0&2&\ldots&0&0&0\cr 0&N-1&0&\ldots&0&0&0\cr\vdots&\vdots&\vdots&\ddots&\vdots&\vdots&\vdots\cr0&0&0&\ldots&0&N-1&0\cr 0&0&0&\ldots&2&0&N\cr 0&0&0&\ldots&0&1&0\cr}.
\label{eq:mat}
\ee
The $N+1$ eigenvalues $\lambda$ and right eigenvectors $r[n]$ of this matrix can be obtained from a generating function
\be
f(u,v)=\sum_{n=0}^N u^n v^{N-n} r[n].
\label{eq:diff}
\ee
When used with the eigenvalue equation $\sum_{n'} M[n,n'] r[n'] = \lambda r[n]$, $f$ is seen to obey the differential equation
\be
\lambda f = [u {\partial\over \partial v} + v {\partial\over \partial u} ]f,
\ee
whose solution of the required form $f(u,v)=v^N h(u/v)$ is
\be
f_\lambda= K_\lambda~ (v+u)^{(N+\lambda)/2} (v-u)^{(N-\lambda)/2},
\label{eq:gen}
\ee
where $K_\lambda$ is independent of $u$ and $v$.  The series in $u$ and $v$ must terminate, requiring the two exponents in Eq.~(\ref{eq:gen}) to be non-negative integers, and thus implying that the eigenvalues are $\pm N. \pm (N-2), \ldots , \pm 1~ ({\rm{for}~\rm{odd}}~N)~ {\rm{or} ~0~(\rm{for}~\rm{even}}~N)$.


The left (dual) eigenvectors $l[n]$ are generated by
\be
g(u,v)=\sum_{n=0}^N u^n v^{N-n}\pmatrix{N\cr n\cr} l[n],
\label{eq:gen'}
\ee
the coefficient being the combinatorial coefficient defined in Eq.~({\ref{eq:exact}}).  When used with the eigenvalue equation $\sum_{n'} l[n'] M[n',n] = \lambda l[n]$, $g$ is seen to obey the identical differential equation as $ f$, namely Eq.~(\ref{eq:diff}).  The constants in the  solution Eq.~(\ref {eq:gen}) determine  normalization.  One choice is to take $K_\lambda={1\over 2^N}$ for $f_\lambda$ and 
\be
K_\lambda=  \pmatrix{N\cr {N+\lambda\over 2}\cr}
\label{eq:norm}
\ee
for $g_\lambda$, whereupon $r_N [n]$ is given by $P_{eq}[n]$, Eq.~({\ref{eq:exact}}), the stationary normalized solution of Eq.~(\ref{eq:mark}), and $l_N [n] = 1$ for every $n$.   That this choice also achieves the completeness relation for orthonormal eigenvectors,
\be
\sum_\lambda r_\lambda [n] l_{\lambda } [n'] = \delta _{n, n '},
\label{eq:comp}
\ee
can be seen from Eqs.~(\ref{eq:diff}), (\ref{eq:gen}), and (\ref{eq:gen'}).

These considerations facilitate analysis of the approach to equilibrium. The initial condition of flealess Anik may be written using Eq.~(\ref{eq:comp}) as
\be
P_0[n] = \delta_{n,0} = \sum_\lambda r_\lambda [n] l_\lambda [0] ,
\ee
whereupon $t$ steps of the evolution  Eq.~(\ref{eq:mark}) yield
\be
P_t[n] = \big({M\over N}\big)^t \sum_\lambda r_\lambda [n] l_\lambda [0] =\sum_\lambda \big ({\lambda\over N}\big )^t r_\lambda [n] l_\lambda [0].
\ee

The moments of this evolved distribution may now be calculated.   Comparing partial derivatives with respect to $u$ of Eqs.~(\ref{eq:diff})  and (\ref{eq:gen}) one finds
\be
\sum_n n~r_\lambda [n]= {N\over 2} \delta_{\lambda , N} - {1\over 2} \delta_{\lambda , N-2}
\label{eq:mean}
\ee
and
\beqa
\sum_n n(n-1)~r_\lambda [n]&=& {N(N-1)\over 4} \delta_{\lambda , N} - {(N-1)\over 4} \delta_{\lambda , N - 2}\cr &+& {1\over 2}  \delta_{\lambda , N - 4}.
\label{eq:nsq}
\eeqa
Since $l_\lambda [0]$ can be seen to be equal to the $K_\lambda$ of Eq.~(\ref{eq:norm}), one deduces
that
\beqa
\sum_n n~P_t[n] \equiv N  \mu (t)& = &{N\over 2} \big[ 1 - ( 1 - {2\over N})^t \big]\cr &\Rightarrow&{N\over 2} \big[ 1 - {\rm e}^{-2t/N}\big ]
\label{eq:appr}
\eeqa
showing that the mean number of fleas approaches equal partitioning exponentially with an equilibration time $N/2$. The decay as $( 1 - {2\over N})^t = \lambda_2^t$ is actually a general result: in any Markov process the equilibration is controlled asymptotically by the second largest eigenvalue  $\lambda_2$.

In a similar way, it is seen using Eqs.~(\ref{eq:mean}) and (\ref{eq:nsq}) that the mean square fluctuation of the number at time step $t$ is given, within the exponential approximation of the last line of  Eq.~(\ref{eq:appr}),  by
\be
\sum_n (n -\mu (t) )^2~P_t[n] = N \mu (t)(1 -\mu (t)).
\ee
This shows, interestingly, that the relation between the mean and width of a binomial distribution for the probabilities associated with tossing a biased coin is preserved during stages of the evolution long before equilibrium is reached.

These methods also permit the exact calculation of the integrated autocorrelation time $\tau_A$, defined in Eqs.~(\ref{eq:corr}, \ref{eq:auto}),  for this simple model.  As an example, we consider the number $n$ of fleas on Anik and calculate the corresponding autocorrelation time $\tau_n$. We need to calculate the average $\lan n^\prime n\ran$, where $n^\prime$ is the number of fleas a given number of hops later than an $n$-flea state.  For $t$ hops, this average is
\be
C_t\equiv\sum_{n. n^\prime} n^\prime {M^t [n^\prime, n]\over N^t} n P_{eq } [ n],
\label{eq:thops}
\ee
where $M$ is given in Eq.~(\ref{eq:mat}) and $P_{eq}$ is the equilibrium distribution of Eq.~(\ref{eq:exact}).  In Eq.~(\ref{eq:thops}), $n$ is picked at random from the known correct distribution and $n^\prime$ is correlated with $n$ via the conditional probability for $t$ hops.

Now $M$ can be represented in terms of its eigenvalues and eigenvectors as
\be
M[n^\prime, n] = \sum_\lambda r_\lambda [n^\prime]~ \lambda~ l_{\lambda}[n],
\ee
and it folows, using the orthonormality relation $\sum_n l_\lambda [n]~ r_{\lambda^\prime}[n] = \delta_{\lambda,\lambda^\prime}$, that
\be
{M^t[n^\prime, n]\over N^t} = \sum_\lambda r_\lambda~[n^\prime]\big( {\lambda\over N}\big)^t~ l_{\lambda}[n].
\ee
The contribution of the highest eigenvalue $\lambda = N$ to this sum, obtained from the eigenvectors given above Eq.~(\ref{eq:comp}), is found to be independent of $n$ and equal to $P_{eq} [n^\prime ]$ for any $t$.  This convenient fact leads to the identity
\be
{M^t[n^\prime, n]\over N^t} - P_{eq}[{n^\prime]= \sum_{\lambda\not= N}r_\lambda~[n^\prime]} \big({\lambda\over N}\big)^t~ l_{\lambda}[n].
\ee
When the right hand side of this form is substituted into Eq.~(\ref{eq:thops}) one encounters the average given in Eq.~(\ref{eq:mean}), and also the average
\be
\sum_n  n~ P_{eq }[n]~ l_{\lambda}[n] = \pmatrix{N\cr {N+\lambda\over 2}\cr} \big[ {N\over 2} \delta_{\lambda , N} - {1\over 2} \delta_{\lambda , N-2} ],
\ee
which has been evaluated via a partial derivative with respect to $u$ of the generating function $g$ in Eq.~(\ref{eq:gen}).  Thus only the eigenvalue $\lambda = N - 2$ contributes to the happily simple result
\beqa
\sum_{t=1}^\infty [C_t -\lan n\ran^2] &=&  \sum_{\lambda\not= N}\sum_{n,n^\prime} n^\prime r_\lambda[n^\prime] { {\lambda / N}\over 1 -  {\lambda / N}}~ l_{\lambda}[n] n P_{eq }[n]\cr &=& 
\pmatrix{N\cr {N-1}\cr}{1 -2/N \over 2/N} ({1\over 2})^2 \cr &=& {N ( N -2)\over 8}.
\eeqa
Since the equilibrium variance of $n$ is $N/4$, we see by comparison with Eq.~(\ref{eq:auto}) that the integrated autocorrelation time for sampling $N$ fleas one at a time is
\be
\tau_n (N) = {N - 2\over 2},
\ee
so that $M$ single hops are equivalent to only
\be 
M_{eff} = {M\over 2\tau_n + 1} = {M\over N - 1}
\ee
trials of N-flea configurations.  In our simulation of $N=50$ fleas we determined $\tau_n=24.0$ using the binning analysis, in perfect agreement with the prediction $\tau_n = (N-2)/2=24.$

It is gratifying that the above learned considerations show that randomizing the number of `heads' among $N$ coins by arbitrarily choosing and turning over $1$ is $1/(N-1)$ times as effective as tossing all $N$ at once.  As mentioned in the Introduction, it is the less efficient processes which are typically at  work in physical situations.


\begin{thebibliography}{99}
\bibitem{LasVegas} If the method had been invented only recently the name would most likely have been ``Las Vegas method''.
\bibitem{ODD}An odd number $N_{\rm hop}$ is important since $n$ is even after and even number of flea hops and odd after an odd number. Choosing an even value of $N_{\rm hop}$ we would record only even entries for $P[n]$.
\bibitem{RC} In the appendix we explicitly calculate the autocorrelation function for this model and show that it indeed decays exponentially..
\end{thebibliography}
 \enddocument